\documentclass[%
 aip,
 amsmath,amssymb,
 reprint,%
]{revtex4-1}

\usepackage{siunitx}
\usepackage{graphicx}
\usepackage{dcolumn}
\usepackage{bm}

\usepackage[utf8]{inputenc}
\usepackage[T1]{fontenc}
\usepackage{mathptmx}
\usepackage{etoolbox}
\usepackage{gensymb}
\usepackage{braket}

\makeatletter
\def\@email#1#2{%
 \endgroup
 \patchcmd{\titleblock@produce}
  {\frontmatter@RRAPformat}
  {\frontmatter@RRAPformat{\produce@RRAP{*#1\href{mailto:#2}{#2}}}\frontmatter@RRAPformat}
  {}{}
}%
\makeatother
\begin{document}

\preprint{AIP/123-QED}

\title[All-in-one Quantum Diamond Microscope]{All-in-one Quantum Diamond Microscope for Sensor Characterization.}

\author{Connor Roncaioli}
\email{connor.a.roncaioli.civ@army.mil} 
 \affiliation{DEVCOM Army Research Laboratory}
 \affiliation{Quantum Technology Center, University of Maryland}
\author{Connor Hart}
 \email{chart@umd.edu}
 \affiliation{Quantum Technology Center, University of Maryland}
 \author{Ronald Walsworth}
 \email{walsworth@umd.edu}
 \affiliation{Quantum Technology Center, University of Maryland}
\author{Donald P. Fahey}
 \email{donaldpfahey@gmail.com}
 \affiliation{DEVCOM Army Research Laboratory}
\affiliation{Quantum Technology Center, University of Maryland}

\date{\today}


\begin{abstract}
Nitrogen-vacancy (NV) centers in diamond are a leading modality for magnetic sensing and imaging under ambient conditions. However, these sensors suffer from degraded performance due to paramagnetic impurities and regions of stress in the diamond crystal lattice. This work demonstrates a quantum diamond microscope (QDM) for the simultaneous mapping and spatial correlation of key properties of a millimeter-scale NV-diamond sensor chip, including: NV ensemble photoluminescence (PL) amplitude, spin-lattice relaxation time (T$_1$), and homogeneous and inhomogeneous spin coherence lifetimes (T$_2$ and T$_2^*$), as well as lattice stress/strain, birefringence magnitude, and birefringence angle of the diamond crystal.
\end{abstract}

\maketitle

\section{\label{sec:Intro}Introduction\protect}

Solid state quantum sensors, such as the nitrogen-vacancy (NV) center in diamond, are undergoing rapid research and commercial expansion. Example NV-diamond measurement modalities include compact scalar\cite{Taylor2008,Rondin_2014,Eisenach2021,Fescenko2020,Ziem2019, Aslam2023} and vector\cite{Schloss2018,Wang2015} magnetic field sensing and imaging, stable clock oscillators\cite{Hodges2013,MattPolariton}, electric field sensors\cite{Iwasaki2017, Scholten2021}, nanodiamond temperature sensors\cite{Neumann2013, Aslam2023}, mapping of strain in the diamond crystal\cite{Kehayias2019, Barfuss2019}, microwave mode cooling\cite{Zhang2022,FaheyCooling}, extreme environment magnetometry\cite{Toyli2012}, and pressure sensing.\cite{Bhattacharyya2024}

A widely used platform is the quantum diamond microscope (QDM), which employs a dense surface layer of NVs on a macroscopic diamond plate to enable wide-field micron-scale vector imaging of magnetic fields at very low standoff distances.\cite{Levine2019, Barry2020} Useful operation of the QDM requires detailed characterization of NV-diamond sensor properties for particular use cases\cite{Alghannam2019,Osterkamp2019,Ofori-Okai2012}, as well as to identify imperfections that can limit QDM performance.\cite{Kehayias2019,Marshall2021,Knauer2020} Here we demonstrate an "all-in-one QDM" that can image several parameters relevant to NV ensemble sensing across a millimeter diamond plate, including NV spin resonance lineshift, decoherence, depolarization time scales, stress/strain, and birefringence. Measurements are co-registered to simplify analysis of complex features that appear across modalities.

\begin{figure*}
\includegraphics[width=7in]{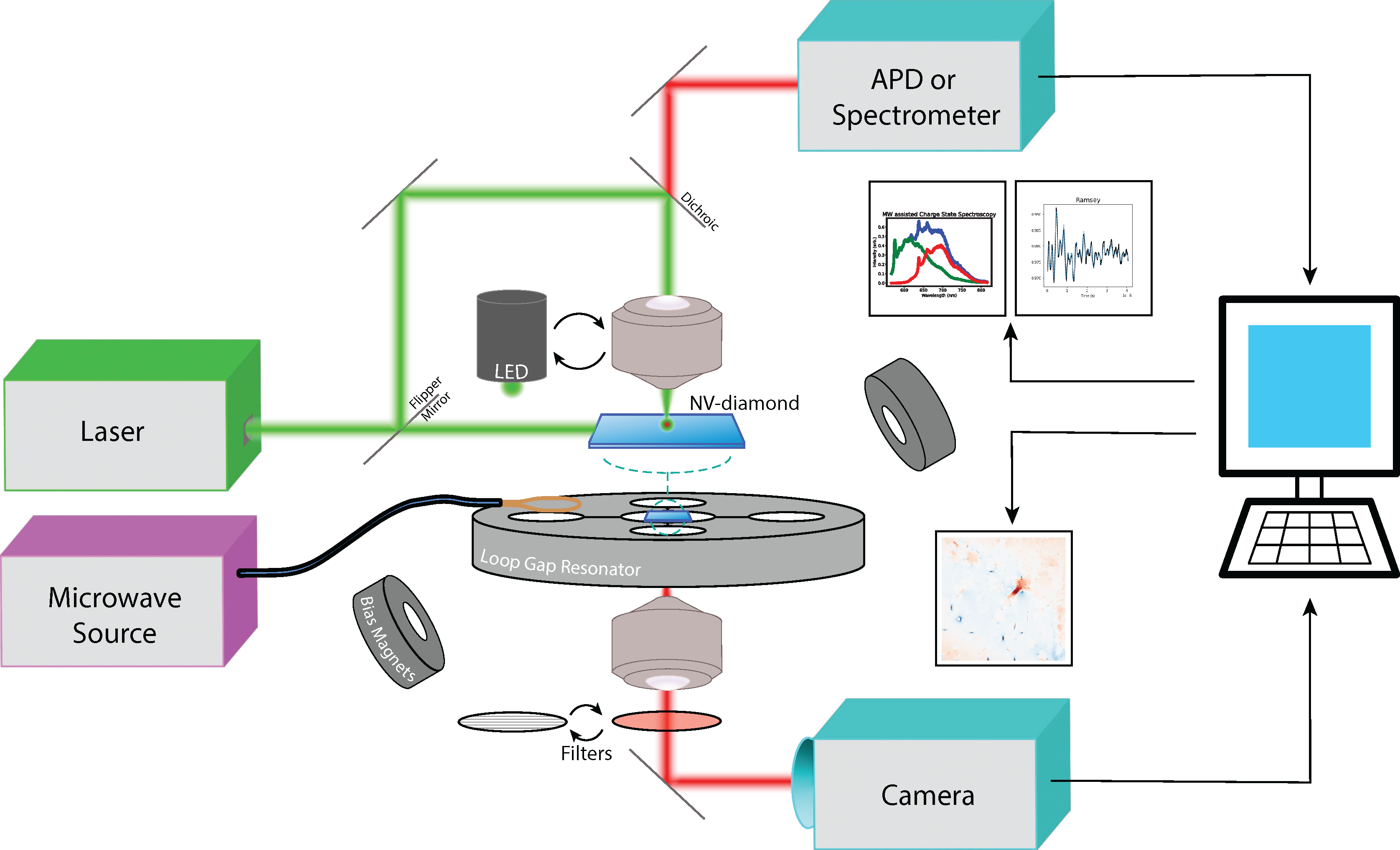}
\caption{\label{figure:Schematic} Schematic of all-in-one QDM for NV-diamond sensor characterization. Two optical excitation pathways are exchanged by use of a flipper mirror, with an epifluorescent option and a non-focused full-sample excitation option. Microwaves (MWs) are coupled in to an edge inductive gap of the loop gap resonator (LGR), creating a strong, homogeneous MW field across the full NV-diamond sample. Permanent magnets create a uniform bias magnetic field that Zeeman shifts NV spin resonances in order to isolate or overlap resonance frequencies as needed. Optical objectives above and below the sample are used for epifluorescent or full field-of-view measurements. The avalanche photodiode (APD) and optical spectrometer are coupled to the above-sample collection pathway. The APD is used for time-resolved NV photoluminescence (PL) measurements, while the spectrometer is used for MW assisted NV charge state measurements. The below-sample objective is coupled to a long-pass filter and camera used to image PL for spatially resolved NV characterization. An LED with linear polarization control and appropriate circular polarization filter can be switched for the top objective and bottom filters, respectively, to create a transmission measurement of optical birefringence using the same below-sample objective and camera.}
\end{figure*}

\section{\label{sec:InstrumentationDesign}Instrument Overview\protect}

The all-in-one QDM is configurable between multiple measurement modalities, as listed in Table \ref{table:modes}. A high-level depiction of the apparatus is illustrated in Fig. \ref{figure:Schematic} and the supplemental material includes a summary of each measurement modality and list of components for each subsystem.

Laser light for NV excitation can be sent through a top objective for point-excitation, or to a periscope that illuminates a mm-scale region of the diamond along a shallow angle. Similarly there are two collection pathways for NV photoluminescence (PL), the top objective can be used to collect epifluorescence and spatially filter it before being sent to either a photodiode for time resolved confocal measurement or a fiber-coupled spectrometer for spectroscopy; or the PL can be collected by an imaging objective mounted beneath the sample space, which directs light to a camera for wide-field imaging. This combination of options affords great flexibility: e.g., the all-in-one QDM can collect and correlate confocal and imaged NV measurements. The top objective can be exchanged with a \SI{530}{\nano\meter} LED with 360\degree linear polarization control, and the filters in the imaging path can be switched with a circular polarization analyzer to change the system into an optical birefringence transmission measurement, as detailed in Section \ref{sec:Birefringence}.

The system is capable of strong MW driving of NV spin transitions (>\SI{1}{\mega\hertz} Rabi frequency) across the full field of view of the imaging objective ($\approx$\SI{5}{\milli\meter}). Achieving an NV Rabi frequency $\approx \frac{1}{\textrm{T}_2^*}$ across the entire diamond sample is necessary to enable simultaneous imaging of key properties such as the NV ensemble spectrum of optically detected magnetic resonance (ODMR). This MW capability is enabled by the use of a metallic loop gap resonator (LGR) innovated at MIT-LL.\cite{Eisenach2018} Details are provided in the supplemental material.

\begin{figure}
\includegraphics[width=3.5in]{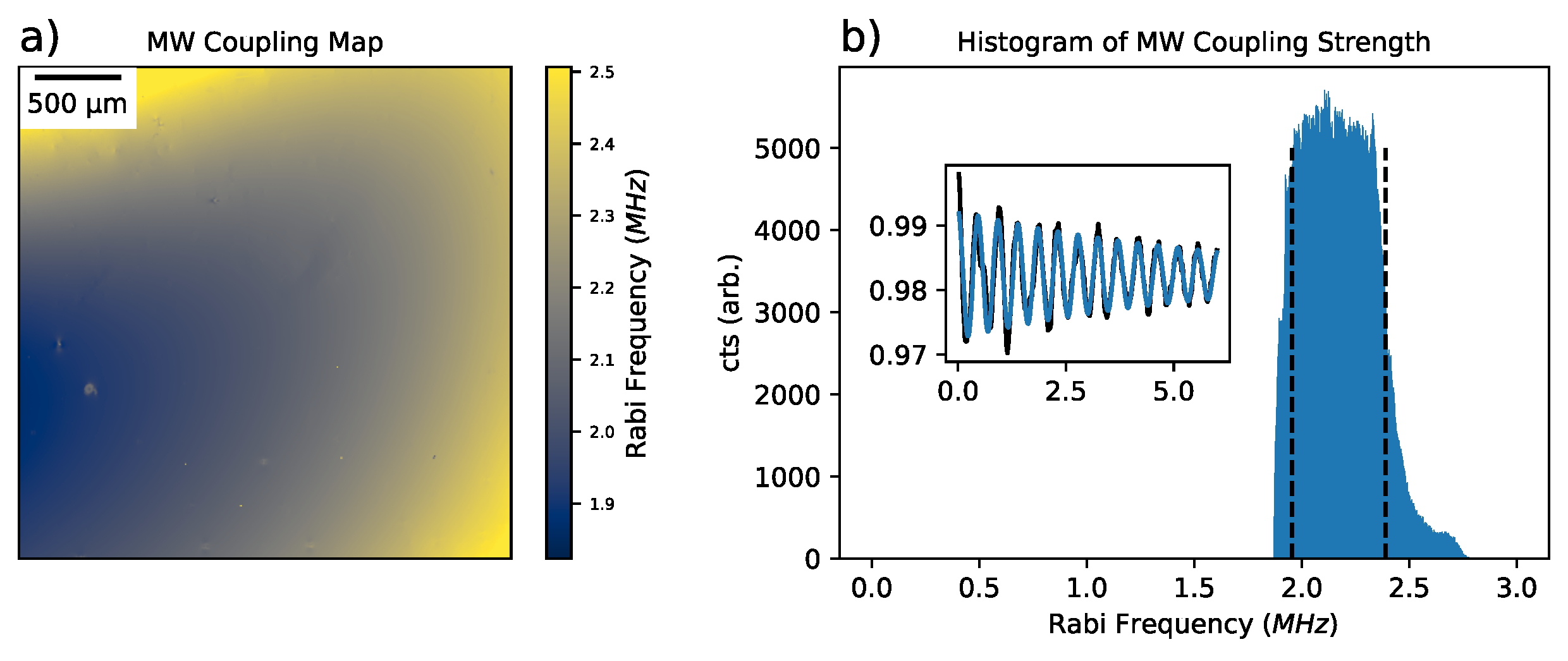}
\caption{\label{figure:RabiCoupling} a) Example map of NV ensemble Rabi frequency produced by all-in-one QDM across $\approx$3x\SI{3}{\milli\meter} region of a diamond sensor plate. NV-diamond sample is described in section \ref{sec:DiamondCaseStudy}. Observed variation in Rabi frequency across field-of-view results from spatial gradients in MW coupling strength to NV spin transitions, e.g., due to inhomogeneous MW drive amplitude or angle of incidence relative to the interrogated NV dipole, or any cause of NV spin resonance lineshift or broadening.  Spatial scale bar is shown. b) Histogram of fit values provides an estimate of Rabi frequency distribution across all-in-one QDM field-of-view. 10th and 90th percentiles, marked with dashed black lines, fall within $\pm$10\% of mean Rabi frequency value. Inset: example single pixel Rabi oscillation data set and associated fit in blue.
}
\end{figure}

\begin{table*}
\caption{\label{table:modes}All-in-one QDM measurement modalities: CLE - confocal laser excitation using epifluorescent illumination, FSLE - full-sample laser excitation, LED - light-emitting diode, PD - photodiode, CMOS - complementary metal-oxide semiconductor camera.}
\begin{ruledtabular}
\begin{tabular}{c|c|c|c}
&\multicolumn{2}{c}{Measurement Details}\\ \hline \hline
\hspace*{1.75cm}Measurement\hspace*{1.75cm}&Pump Geometry\hspace*{0.3cm}&Read-Out Configuration\hspace*{0.3cm}&Purpose\\ \hline
Constant wave-optically detected&CLE&PD&MW resonance frequencies\\ 
magnetic resonance (CW-ODMR)&&&of NV spin transitions\\
Rabi&CLE&PD&MW driving rate of NV spins\\
T$_1$ Relaxometry&CLE&PD&NV spin-lattice relaxation time (T$_1$)\\
Ramsey&CLE&PD&NV spin dephasing time (T$_2^*$)\\
Hahn echo&CLE&PD&NV spin decoherence time (T$_2$)\\
Dynamic decoupling&CLE&PD&Performance of noise-cancelling\\ &&&pulse sequences outside of\\ &&&specified frequency window\\
Photoluminescence (PL) distribution&FSLE or CLE&CMOS&NV PL intensity and epi-spot location\footnote{A supplement to photodiode measurements, showing the precise location of the pump spot relative to the sample in imaged measurements}\\
Imaged CW-ODMR&FSLE&CMOS&NV spin resonance across sample\\
Imaged Rabi&FSLE&CMOS&MW driving rate of NVs across sample\\
Imaged T$_1$ relaxometry&FSLE&CMOS&NV T$_1$ across sample\\
Imaged Ramsey&FSLE&CMOS&NV T$_2^*$ across sample\\
Imaged Hahn echo&FSLE&CMOS&NV T$_2$ across sample\\
Imaged dynamic decoupling&FSLE&CMOS&Performance of noise-cancelling\\ &&&pulse sequences across sample\\
\hline \hline
&Illumination Source\hspace*{.2cm}&&\\ \hline
Imaged quantitative birefringence&LED&CMOS&Spatial distribution of \\ &&&stress/strain within diamond plate\\
\end{tabular}
\end{ruledtabular}
\end{table*}

In order to produce spatially resolved maps of parameters of interest, with micron-scale resolution across a mm-scale field-of-view, upwards of 1 million distinct data sets require fitting. This analysis requires the use of parallel GPU-based fitting software, such as GPUfit and the associated python-based pyGPUfit module.\cite{GPUfit} For CW-ODMR, Rabi, and T$_1$ relaxometry, measurements are reported as contrast:

\begin{equation}
\textrm{contrast} = \frac{\textrm{I}_{\textrm{MWon}}}{\textrm{I}_{\textrm{MWoff}}}
\label{eq:contrast}
\end{equation}

\noindent Here, I$_{\textrm{MWon}}$ is the measured PL signal intensity with MWs on and I$_{\textrm{MWoff}}$ is the reference PL intensity measurement with MWs off. For phase-sensitive modalities (Ramsey and Hahn echo) values are reported as visibility:

\begin{equation}
\textrm{visibility} = \frac{\textrm{I}_{\textrm{MW+}} - \textrm{I}_{\textrm{MW-}}}{\textrm{I}_{\textrm{MW+}} + \textrm{I}_{\textrm{MW-}}}
\label{eq:visibility}
\end{equation}

\noindent Here, I$_{\textrm{MW}+}$ is the measured PL intensity where all MW pulses applied to the NVs have the same phase, and I$_{\textrm{MW}-}$ is PL intensity with the final $\frac{\pi}{2}$ MW pulse of opposite phase. More details, such as the functional form of the various fit procedures and techniques for evaluating fit appropriateness are provided in the supplemental material.

\section{\label{sec:ExperimentalModalities}Experimental Modalities\protect}

\begin{figure}
\includegraphics[width=3in]{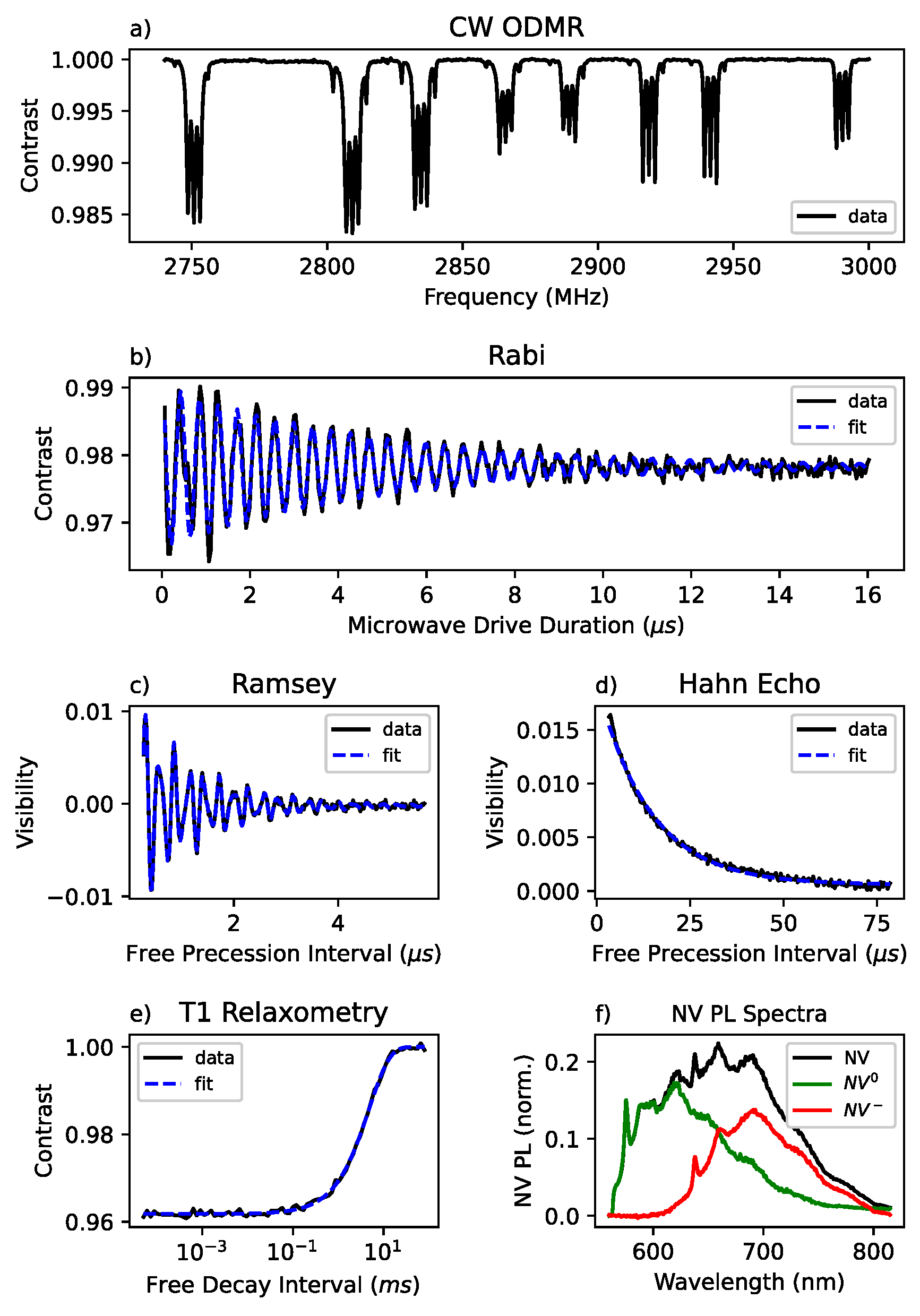}
\caption{\label{figure:Epi}Example NV ensemble measurements for diamond sample described in section \ref{sec:DiamondCaseStudy}, and associated fits for data, collected with all-in-one QDM in epifluorescent configuration. a) CW-ODMR spectra; b) Rabi oscillation; c) Ramsey measurement; d) Hahn echo; e) T$_1$ relaxation; and f) NV PL spectra demonstrating MW-assisted charge state decomposition.\cite{Craik2020} For the CW-ODMR data, LGR resonant frequency $\approx$ \SI{2785}{\mega\hertz} and a bandwidth $\approx$ \SI{80}{\mega\hertz}, which enhances MW driving of nearby resonant NV spectral features arising from different projections of bias magnetic field on four NV orientations within the diamond crystal.}
\end{figure}

\subsection{\label{sec:CWODMR}CW-ODMR\protect}

Confocal continuous wave (CW)-ODMR is used for high-sensitivity measurement of NV spin resonance frequencies at a near-diffraction-limited point and focal depth. Imaged CW-ODMR, performed via non-focused full-sample laser illumination, allows measurement of the NV spin resonance spectrum on a pixel-by-pixel basis, but necessarily collects NV PL from all depths of field. This modality can map interactions that affect the NV spin Hamiltonian, including stress within the diamond. Based on previous work \cite{Kehayias2019,Barfuss2019} we describe the NV stress tensor:

\begin{eqnarray}
    \sigma_{diag} = \frac{1}{4a_1}[M_{z,1}+M_{z,2}+M_{z,3}+M_{z,4}]\\
    \sigma_{XY} = \frac{1}{8a_2}[M_{z,1}+M_{z,2}-M_{z,3}-M_{z,4}]\\
    \sigma_{XZ} = \frac{1}{8a_2}[M_{z,1}-M_{z,2}+M_{z,3}-M_{z,4}]\\
    \sigma_{YZ} = \frac{1}{8a_2}[M_{z,1}-M_{z,2}-M_{z,3}+M_{z,4}]
\end{eqnarray}

where $\sigma_{diag} = \sigma_{XX} + \sigma_{YY} + \sigma_{ZZ}$ is the diagonal stress tensor component, the individual values of which are obscured in present QDM measurements; M$_{z,i}$ are the asymmetric lineshift values for each NV orientation classes within the diamond crystal about their mean value; and $\{\textrm{a}_1,\textrm{a}_2\}=\{\textrm{4.86},\textrm{-3.7}\}$ $\frac{\SI{}{\mega\hertz}}{\SI{}{\giga\pascal}}$ is the spin-stress coupling constant. Using these relationships we can partially reconstruct the stress tensor, including the shear components, from QDM measurements (see supplemental material). Note that determination of the individual uniaxial terms ($\sigma_{XX}$, etc.) is possible by changing the bias magnetic field direction, though this is not implemented here.\cite{Kehayias2019}

\begin{figure*}
\includegraphics[width=7in]{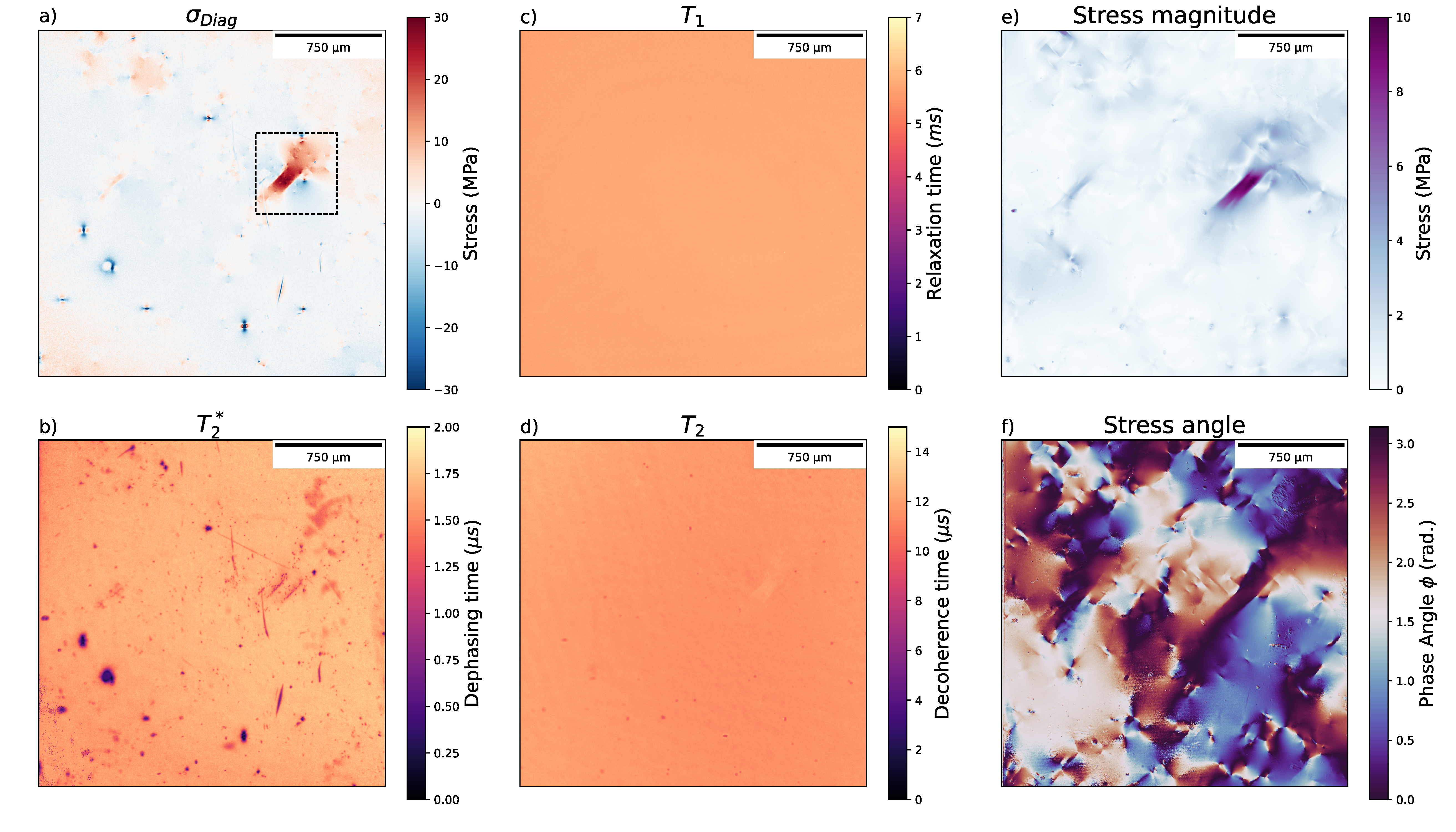}
\caption{\label{figure:FullDiamond} Example of NV-diamond sample characterization performed with all-in-one QDM and displayed as image data. a) $\sigma_{diag}$ extracted from CW-ODMR; b) T$_2^*$ from Ramsey; c) T$_1$ from relaxometry; d) T$_2$ from Hahn echo; e) stress magnitude and f) stress angle determined from optical birefringence. Correlations in a)-f) relate to regions of high stress and stress gradients.
}
\end{figure*}

\begin{figure*}
\includegraphics[width=7in]{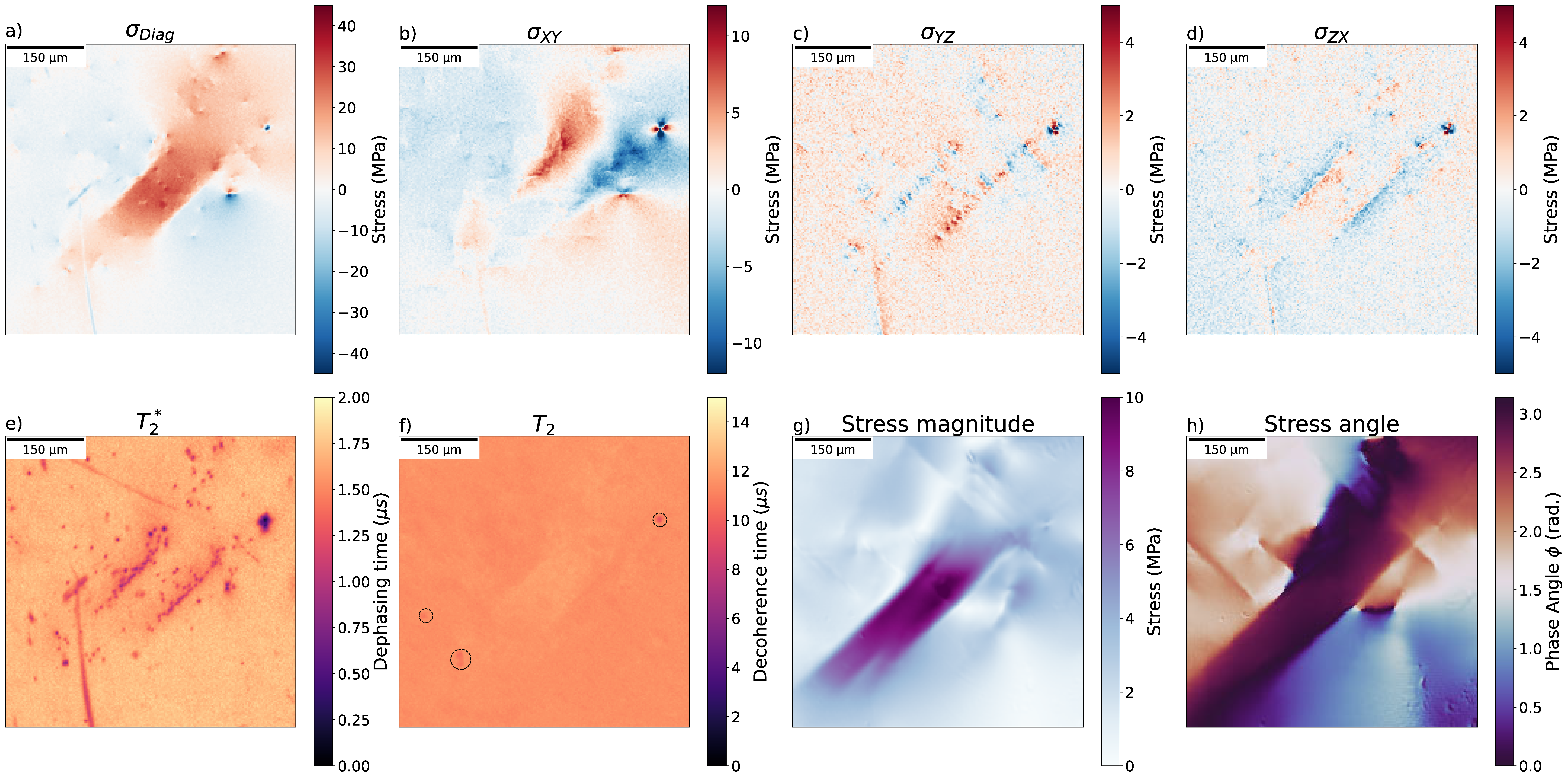}
\caption{\label{figure:ZoomDiamond} Expanded-view QDM characterization images for region of NV-diamond sample indicated by box in Fig. \ref{figure:FullDiamond}a): this  region of enhanced stress appears as a channel between two highly localized linear regions. a) Diagonal stress tensor component $\sigma_{diag}$ extracted from CW-ODMR corresponds to isometric stress. b-d) Off diagonal stress tensor components correspond to shear stress, with $\sigma_{XY}$ being in-plane shear and $\sigma_{YZ}$ and $\sigma_{ZX}$ being out-of-plane shear components. e) $\textrm{T}_2^*$ map shows qualitative correlation with stress measurements. f) $\textrm{T}_2$ map is near-uniform with the exception of a few marked regions. g) stress magnitude and h) stress angle determined from optical birefringence measurements.
}
\end{figure*}

\subsection{\label{sec:PulsedProtocols}Pulsed Protocols\protect}


Pulsed NV measurements are perfomed in either the confocal or wide-field imaging modes, including protocols for Rabi oscillation, Ramsey, Hahn echo, and $\textrm{T}_1$ relaxometry, see Figs. \ref{figure:Epi} and \ref{figure:FullDiamond}. Importantly, the all-in-one QDM allows such measurements over a millimeter-scale field-of-view — whereas past measurements were limited to $\approx$\SI{100}{\micro\meter}\cite{Levine2019} — due to integration of the LGR to provide strong, homogeneous MW driving of NV spin transitions.

Imaging the NV Rabi frequency allows evaluation of MW field homogeneity. Ramsey and Hahn echo measurements are used to evaluate the NV ensemble spin dephasing time (T$_2^*$); and are conducted at a fixed MW detuning ($\approx$ \SI{5}{\mega\hertz}) from the NV center frequency in order to have sufficiently fast oscillation to properly fit the decay envelope. Hahn echo measurements are used to measure the NV ensemble decoherence time (T$_2$). Both Ramsey and Hahn echo measurements are collected with phase dependence of the final MW pulse in order to measure visibility instead of contrast for improved noise cancellation. Finally T$_1$ relaxometry provides an upper limit to NV spin lifetimes in the sample. Dynamical decoupling schemes are feasible for this system but are not implemented in the present work. Examples of imaged fitting parameters extracted from these measurements are provided in Figs. \ref{figure:FullDiamond} and \ref{figure:ZoomDiamond}. Correlation is observed between high-contrast features in the maps of stress and T$_2^*$, which is consistent with past studies of CW-ODMR and Ramsey microscopy of NV-diamond samples.\cite{Marshall2021} However, the T$_2$ and T$_1$ maps are largely homogeneous, as anticipated for the high-quality NV-diamond sample studied here. More detail is provided in section \ref{sec:DiamondCaseStudy}.

\subsection{\label{sec:Birefringence} Optical Birefringence and Stress Imaging\protect}
The cubic symmetry of ideal diamond has zero optical birefringence; however, real diamond samples typically have local stress that breaks this symmetry, resulting in observable birefringence.\cite{Lang1967} The all-in-one QDM provides a quantitative measure of stress in the diamond sample via a switchable stage that brings an LED (wavelength = $\sigma$ = \SI{530}{\nano\meter}) with a rotating linear polarizer above the diamond sample and a circular analyzer after the below-sample objective. In this operational mode, optical transmission measurements yield the full 2D anisotropic retardance vector without adjusting the sample.\cite{Kehayias2019} The stress angle $\phi$ and retardance phase $\delta = \frac{2\pi}{\lambda} \Delta \textrm{nL}$, where L is the sample thickness and $\Delta n$ is the difference in refractive index between the two polarizations, are related to the optical transmission intensity by the following equation:

\begin{equation}
    \textrm{I}_\textrm{i}=\frac{1}{2}\textrm{I}_\textrm{0}[1 + \textrm{sin} 2(\alpha_\textrm{i} - \phi) \textrm{sin} \delta]
\end{equation}

\noindent By measuring I$_\textrm{i}$ as the linear polarizer angle $\alpha_\textrm{i}$ rotates through 180\degree, $\phi$ and sin $\delta$ are determined. $\delta$ is related to the optical wavelength $\lambda$ and the diamond stress magnitude (|$\sigma$|) and the index of refraction n=2.42 by the equation:

\begin{equation}
|\sigma|\approx\frac{2}{3\pi}\frac{|\delta|\lambda}{\textrm{Ln}^3\textrm{q}_{\textrm{iso}}}
\end{equation}

\noindent where q$_\textrm{iso}$=0.3x10$^{-12}$\SI{}{\pascal}$^{-1}$ is a piezo-optical coefficient.\cite{Howell2010}

Stress determined in this way may be buried deep in the undoped diamond, far away from the active layer of NVs. Additionally, small regions of intense stress in the NV layer may not be detectable by birefringence against the contribution of a large undoped diamond, if the active layer is comparatively small. Furthermore, if the diamond has significant strain or the sample is of sufficient thickness, the optical birefringence measurement can become phase-ambiguous once the difference between the slow and fast axes $|\delta|\ge\frac{\lambda}{2}$. These are all reasons that multi-modal detection is crucial for complete characterization of NV samples. Comparison of measurements of stress determined by CW-ODMR (contribution confined to the NV layer) to optical birefringence (contribution from the entire diamond) are presented in Figs. \ref{figure:FullDiamond}a), \ref{figure:FullDiamond}e), and \ref{figure:FullDiamond}f).

\subsection{\label{sec:Spectroscopy}Optical Spectroscopy and NV Charge State Characterization}

The all-in-one QDM can perform optical spectroscopy from a diamond sample excited using either confocal or shallow-angle excitation. Modulating NV PL spectra using resonant MWs allows determination of the relative concentration of NV charge states (NV$^\textrm{-}$ and NV$^\textrm{0}$).\cite{Craik2020} Note that only the NV$^\textrm{-}$ charge state exhibits spin-state-dependent PL, and hence is useful for quantum sensing.

NV charge states are metastable and dependent on both intrinsic material and extrinsic experimental conditions. Intrinsic conditions that can affect NV charge state include local availability of electron donors by doping\cite{Herbschleb2019} and local chemical potential\cite{Sinan2016}. Some samples can have coexisting regions of different charge state ratios.\cite{Savvin2021} Extrinsic conditions that can affect the NV charge state include electric field bias,\cite{Lozovoi2020}\cite{Lozovoi2023} laser intensity,\cite{MANSON2005} and excitation wavelength.\cite{Aslam_2013} NV$^\textrm{-}$ fraction as a function of laser intensity is particularly important to characterize since it falls off at high intensities, which needs to be balanced against NV PL for optimal sensing.\cite{Barry2020}

\section{\label{sec:DiamondCaseStudy}Diamond Case Study}

As a demonstration of the all-in-one QDM, we characterize a quantum grade NV-diamond sample produced by Element Six. This sample is 3x3x0.5mm, with a \SI{20}{\micro\meter} thick 16ppm N, 3ppm NV layer. The sample is treated with a tri-acid perchlorate solution\cite{AcidWash} in order to remove surface contaminants and characterized by XPS to discern surface composition; additional details are provided in the supplemental material. The sample is then characterized with the all-in-one QDM, performing both confocal and imaged CW-ODMR, Rabi, Ramsey, and Hahn echo, as well as imaged optical birefringence of the entire sample, imaged stress of the NV layer, and confocal MW-assisted spectroscopy to determine relative NV charge states (see Figs. \ref{figure:RabiCoupling}, \ref{figure:Epi}, \ref{figure:FullDiamond} and \ref{figure:ZoomDiamond}). After alignment, the diamond sample remains untouched until all measurements are completed, so that imaged modalities are coregistered to within one pixel, enabling precise correlation of features between measurements.

For the wide-field imaging modalities, significant heating of the diamond sample is caused by laser illumination at typical input power of a few watts. Heating leads to an asymmetric lineshift of NV spin resonances, which can result in a lineshift gradient across the sample. To mitigate the effect of the heat load, the diamond sample is mounted cantilever style to a thermally conductive SiC plate; and a \SI{5}{\micro\second} delay is placed between the laser initialization pulse and the MW protocol used for NV measurements. Cantilever mounting also simplifies the subtraction of the birefringence background, since the only contribution comes from free space and rigidly aligned optics.

Fig. \ref{figure:FullDiamond} presents example images of the diamond samples's NV ensemble asymmetric resonance lineshift in CW-ODMR (corresponding to the isometric stress component $\sigma_{diag}$ within the NV layer), T$_2^*$, T$_1$, and T$_2$; as well as optical birefringence over the full sample thickness, yielding both the relative stress magnitude and angle. Regions of stress typically correspond to NV spin resonance broadening and reduced T$_2$, and T$_2^*$, limiting the sensitivity to magnetic fields in these locations.\cite{Kehayias2019, Barfuss2019} Over 90\% of the sample has measured T$_2^*$ within 10\% of the median value; and almost 99\% of the sample has T$_2$ within 5\% of the median T$_2$ value. Avoiding the few regions of enhanced stress and reduced T$_2^*$ and T$_2$ is thereby enabled by all-in-on QDM characterization.

Figure \ref{figure:ZoomDiamond} shows a specific region of enhanced stress at higher spatial resolution, selected from the QDM images of Fig. \ref{figure:FullDiamond}. In Fig. \ref{figure:ZoomDiamond}a) the isometric stress component ($\sigma_{diag}$) shows a region of increased stress that is between two linear regions in a channel $\approx$\SI{100}{\micro\meter} across. Additional stress petals are visible nearby as much smaller features that are tens of microns across. Fig. \ref{figure:ZoomDiamond}b) shows that the two edges of the region of enhanced stress have related, but opposite, polarity shear components that extend across the region of increased $\sigma_{diag}$. Figs. \ref{figure:ZoomDiamond}c) and \ref{figure:ZoomDiamond}d) demonstrate that the out-of-plane shear components are localized only to the edges of the region of increased stress, and do not extend into the internal region. Figs. \ref{figure:ZoomDiamond}e) and \ref{figure:ZoomDiamond}f) show that increased isometric stress is correlated with (and likely causal to) significantly reduced T$_2^*$, but with minimal effect on T$_2$. In particular, the edges of the stress region, which correspond to localized behavior in the shear components or to steep gradients in isometric stress, correlate with decreased T$_2^*$ values. Figs. \ref{figure:ZoomDiamond}g) and \ref{figure:ZoomDiamond}h) show stress magnitude and phase, respectively, determined from optical birefringence measurements; and provide quanitative confirmation of high stress regions determined from NV measurements.

\section{\label{sec:Conclusion}Conclusion and Future Work\protect}
We present a multi-functional quantum diamond microscope (QDM) for comprehensive characterization of NV-diamond samples. The apparatus employs automated components that enable remote operation. With modest future development, we envision push-button operation of all modalities described in this paper without user intervention. More significant future effort will likely be needed to extend the capability of this "all-in-one QDM" to remote operation on multiple samples in sequence, as a result of challenges with careful sample alignment, background subtraction with regards to birefringence, as well as good thermal heat-sinking.

The apparatus described in this paper should be extendable to any optically driven and detected solid-state quantum defect that operates at room temperature. There is a vast library of potential defects to be explored for optically driven quantum sensing, including vacancies and site defects\cite{Falk2013}, transition metal\cite{Diler2020} and rare earth dopants\cite{Zhang2020}\cite{Spindlberger2019} in both diamond and other semiconductor materials. The scale of characterization demands will continue to grow alongside developments in semiconductor defect systems, underlining the need for rapid characterization systems as described in this paper.

\begin{acknowledgments}
The authors thank Dr. Brenda VanMil, Dr. Sean Blakley and Dr. Matthew Trusheim for insightful conversations on solid state systems and NV physics. We also thank Dr. Adam Biacchi at NIST for implementing the conventional fumehood safe tri-acid diamond cleaning method used in this paper, as well as XPS analysis of the sample surface to confirm a successful cleaning and to interrogate surface chemistry. Support for this work was provided by the U.S. Army Research Laboratory, including under Contract No. W911NF1920181, and the University of Maryland Quantum Technology Center.
\end{acknowledgments}

\appendix

\nocite{*}
\bibliography{aipbib}

\end{document}


\maketitle

\section{Instrument}

\begin{figure}[htbp]
\centering
\fbox{\includegraphics[width=.8\linewidth]{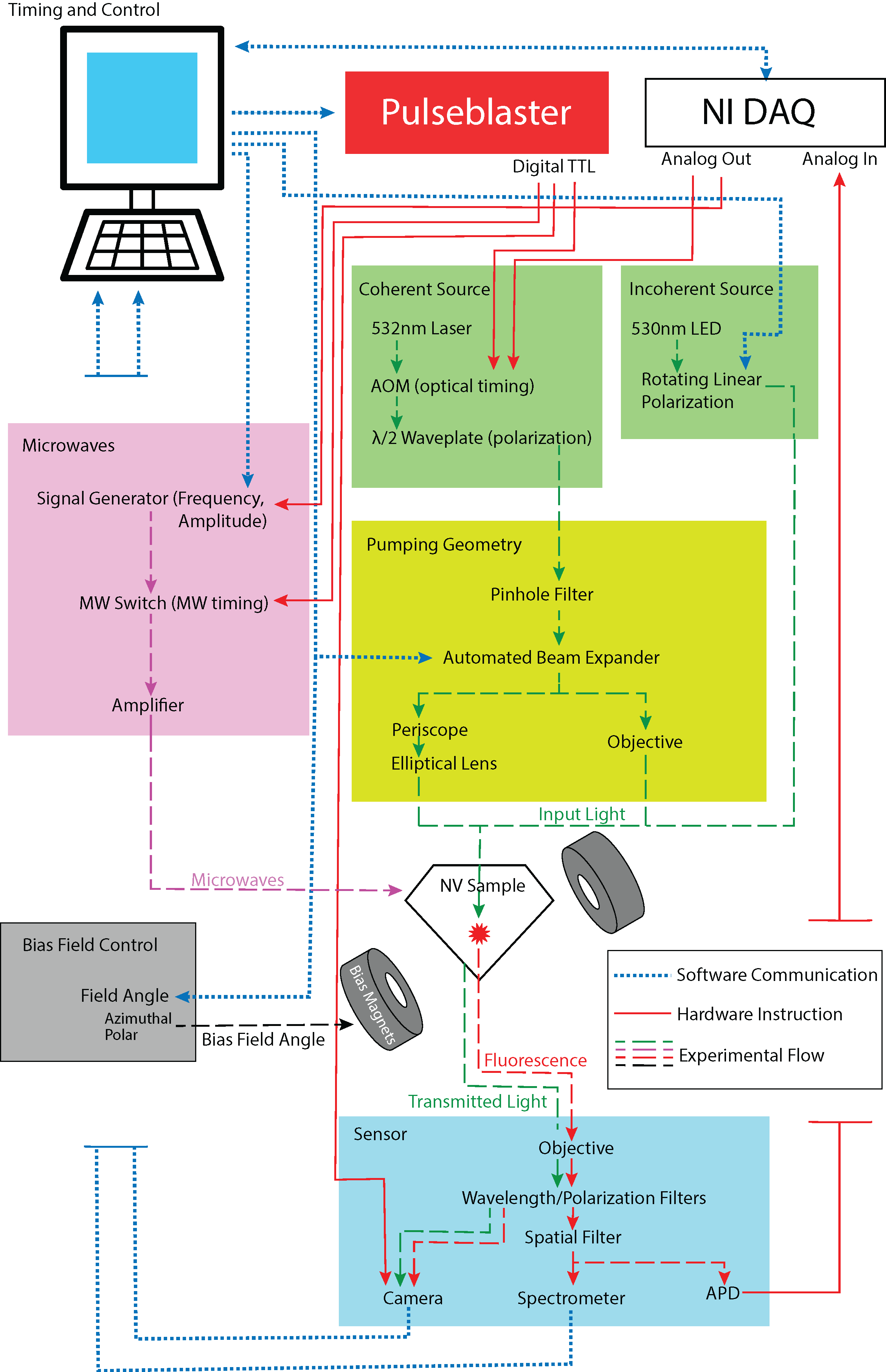}}
\caption{Schematic of all-in-one QDM components and their interrelationships.}
\label{fig:diagram}
\end{figure}

\begin{figure}[htbp]
\centering
\fbox{\includegraphics[width=1\linewidth]{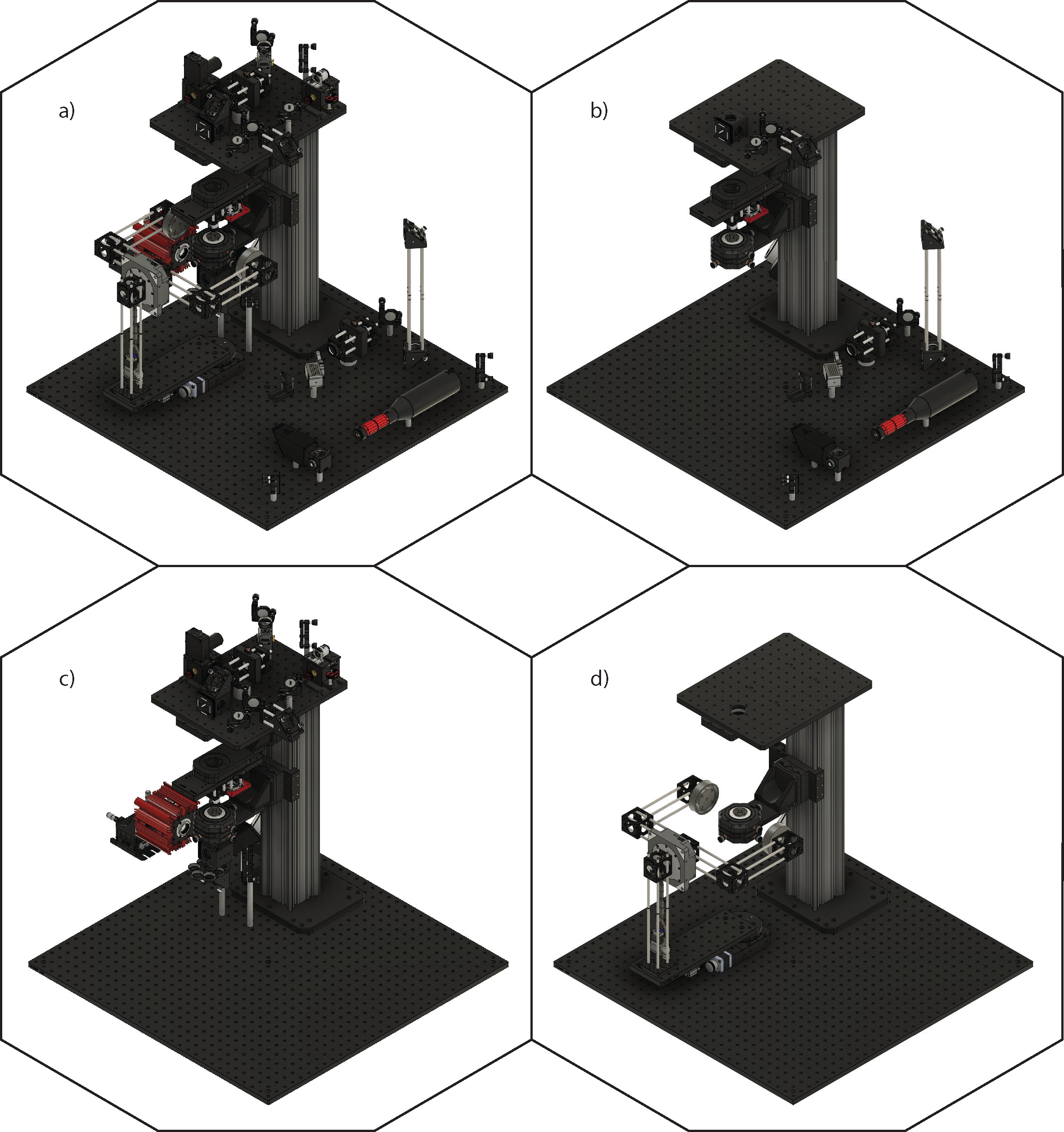}}
\caption{CAD setup of all-in-one QDM subsystems. a) Optical-breadboard components, with the exception of acousto-optic modulator (AOM). Electronics and non-LGR MW components (including antenna coupled to LGR) are omitted. b) Components involved in shaping and polarization of input laser signal, including beam splitter for linear polarization control, motorized beam expander, optional pinhole for TEM00 beam mode shaping, periscope option for low-angle excitation, LED module and rotator on switchable stage, and top objective for transmission birefringence measurement option. c) NV PL and LED transmission collection optics, including long-pass filter, optional pinhole for confocal image, and flipper mirror for PL collection either by photodiode or fiber-coupled spectrometer. d) Bias magnet system, including two motors for control over field angle.}
\label{fig:CAD}
\end{figure}

Fig. \ref{fig:diagram} is a schematic diagram of each component involved in the full apparatus. Fig. \ref{fig:CAD} shows the all-in-one QDM in CAD render.

\subsection{\label{sec:OpticalControl}Optical Control\protect}

A Lighthouse Photonics Sprout DPSS laser (up to \SI{10}{Watt} output power) illuminates NVs for PL measurements. Typical input power to the sample is \SI{80}{\milli\watt} for confocal measurements and \SI{1}{\watt} for wide-field imaging. An acoustic optical modulator (AOM) controls optical switching and intensity of the laser, with $\approx$ \SI{100}{\nano\second} switching time. The optical polarization is controlled by a half-waveplate, a spatial light filter selects the TEM$_{00}$ mode for spatial consistency, and an automated beam magnifier controls the collimated beam spot size, resulting in a collimated zeroth Gaussian mode beam with a controllable beam width and well-defined linear polarization. A beam spot matched to the input objective is ideal for confocal measurements, while a larger beam ($\approx$\SI{5}{\milli\meter}) is necessary for full-sample illumination.

Epi-illumination for point measurements is achieved by directing the laser through the top objective (Olympus PLN 4x) to excite a small ($\approx$ \SI{10}{\micro\meter}) region of the NV layer in the diamond samples. Full-sample illumination for imaging uses a long-focal length cylindrical lens (f=\SI{300}{\milli\meter}) and a periscope that is located next to the microscope body. To evenly illuminate the entire sample, the periscope is designed to bring the beam in at a shallow (<4$\degree$) angle, while a cylindrical lens is used to focus the beam down along the axis perpendicular to the sample face.

Optical birefringence measurements of the diamond sample utilize a green LED (\SI{530}{\nano\meter}) with 360$\degree$ rotatable linear polarization, which swaps with the top objective using a switchable mount. Light transmits through the diamond sample then passes through a circular-polarization analyzer, which sits in a filter turret along the lower imaging pathway, after the below-sample objective. If the analyzer is removed, the LED directly images the sample, which is useful for preliminary alignment and establishing focus on the intended sample surface.

\subsection{\label{sec:BiasFieldControl}Bias field control\protect}

Moving between different sensing modalities requires vector control of the bias magnetic field. With the bias magnetic field along the diamond <111> axis, the spin properties of a single NV orientation can be studied. With the bias magnetic field along the <100> axis, the spin transition frequencies for all four NV orientations overlap and increase contrast. When doing CW-ODMR for all NV orientations, the bias field is set to an intermediate low-symmetry direction that evenly splits the NV spin resonances for the four orientations in order to maximize the spectral dynamic range of each NV orientation class. The QDM uses a $\approx$ 50G magnetic field because a large field is advantageous to increase the spectral dynamic range available. Field strength can be adjusted by symmetrically moving the magnets closer to or further from the diamond sample.

Bias field control uses a system of SmCo magnets mounted on a dual-axis motorized gantry. The gantry provides angular control of about $60\degree$ around the azimuthal and polar angles. This range exceeds the $\approx 55\degree$ required along either axis in order to fully cover the fundamental symmetry group (C$_{3v}$) of diamond, allowing alignment of the bias magnetic field along any and all of the four NV orientations. SmCo bias magnets are used instead of Helmholtz coils to apply a strong, uniform magnetic field due to their thermal stability and independence from DC power fluctuations.  Field homogeneity was <1\% across mm-scale diamond samples, as measured using the CW-ODMR technique and was found to vary by  <1\%, confirming sufficient bias field homogeneity. If further control over the bias field is required, Helmholtz coils are a viable alternative or supplement to permanent magnets.

\subsection{\label{sec:MicrowaveControl}Microwave Control\protect}

A LGR tuned to $\approx$\SI{2.87}{\giga\hertz} delivers a MW signal for NV manipulation, with typical Rabi frequency of a few-MHz (> NV $\frac{1}{T_2^*}$) across a mm-size diamond sample and Q $\approx$ 35 and hence bandwidth $\approx$ \SI{80}{\mega\hertz}. A SRS SG386 signal generator with I/Q control option connected to a fast MW switch and a 16 W amplifier supplies a MW signal in the 2700-\SI{3000}{\mega\hertz} band to the LGR. Phase-dependent pulsed protocols depend on I/Q control, e.g. Ramsey and Hahn echo.

The LGR has a wide central loop about \SI{8}{\milli\meter} across, which both enhances MW signal amplitude and intensity across the cavity.\cite{Eisenach2018} Use of glass as a dielectric shim increases the LGR bandwidth, allowing all NV spin resonances to be addressed, including pulsed measurements across a few-mm diamond sample. MW signals are provided to the LGR using a simple shorted loop antenna inductively coupled to the \SI{5}{\milli\meter} outer loops. In future work, a split ring antenna and a stub tuner could be used  to enhance MW coupling.

\subsection{\label{sec:OpticalCollection}Optical Collection}

There are two primary optical detection pathways, both of which use a 4x NA=0.1 objective with \SI{5}{\milli\meter} focal length for a diffraction-limited resolution of \SI{3}{\micro\meter} at 600 nm (Olympus PLN 4x). In the confocal path, a long-pass dichroic is used to direct the pump laser into the top objective. The top objective also collects NV PL, which passes through the dichroic and into a collection pathway. A spatial filter collects only PL from a region of uniform illumination at the center of the pump region, as well as confocally limits the z-axis collection region. If high signal intensity is preferred, the collection spatial filter can be removed to collect PL from the low-intensity region around the focal point of the objective, as well as from NVs that are out of focus. A flipper mirror then directs the collected PL either to a photodiode for time-resolved PL intensity measurements, or to a spectrometer. 

Below the sample stage is a second objective for wide-field imaging, limited to a field of view of \SI{5}{\milli\meter}. NV PL collected by this objective passes through a \SI{633}{\nano\meter} longpass filter (Semrock LP02-633RE-25), and focuses onto a CMOS camera with a 200 mm tube lens (Basler aca1920-155um). This optical detection pathway is compatible with low-angle ($<4\degree$) illumination of the entire sample; or with pump light provided via the top objective, in which case the exact point of NV excitation can be determined by comparison with the PL location from a mapped image of the sample. To measure optical birefringence of the diamond sample, the dichroic is removed and a circular polarization filter is selected. In this way, polarized green LED light passed through the sample can be imaged without having to realign either the objective or the eyepiece optic, ensuring that NV PL collected from other measurements is co-aligned with the birefringence measurement. 

\subsection{\label{sec:ProgramControlandDataAnalysis}Program Control and Data Analysis\protect}

Equipment timing is controlled by ultrafast digital outputs from a pulseblaster (PBESR-PRO-500-USB-RM). Analog control of devices and analog input from photodiodes is controlled by an National Instruments USB-6363 digital acquisition board (NI-DAQ). The NI-DAQ is timed to the same clock used by the pulseblaster in order to prevent timing drift between the two devices. Operation of the all-in-one QDM is performed using customized python software.

\section{Data Fitting}

Imaged data sets in this work are three-dimensional data cubes. Two array axes encode spatial (X, Y) information, and a third axis is for the dependent variable of a given measurement, typically frequency (CW-ODMR) or characteristic measurement time (Ramsey, Hahn Echo etc.). Finally, the resident float for a given array element is the dependent variable. For CW-ODMR, Rabi, and T1 relaxometry measurements, this is contrast:

\begin{equation}
\textrm{contrast} = \frac{\textrm{I}_{\textrm{MWon}}}{\textrm{I}_{\textrm{MWoff}}}
\label{eq:contrast}
\end{equation}

For phase-sensitive measurements, such as Ramsey and Hahn echo, the dependent variable is visibility:

\begin{equation}
\textrm{visibility} = \frac{\textrm{I}_{\textrm{MW+}} - \textrm{I}_{\textrm{MW-}}}{\textrm{I}_{\textrm{MW+}} + \textrm{I}_{\textrm{MW-}}}
\label{eq:visibility}
\end{equation}

A predictive function is fit to these variables in order to extract a parameter of interest. E.g., T$_2$ is calculated by identifying the characteristic decay time of the Hahn echo visibility. Such fitting is straightforward with conventional analysis packages for a single dataset (i.e., along one axis of the data cube corresponding to data from a given pixel). However fitting rapidly becomes non-trivial for mm-scale images that are on the order of 1000x1000 pixels, corresponding to $\approx 10^6$ data sets. 

T$_2$ is determined by a fit to a simple decaying exponential:

\begin{equation}
\textrm{visibility(T}_2\textrm{)} = \textrm{A}  \textrm{e}^{-\tau \times \kappa(\textrm{T}_2)}
\end{equation}

T$_1$ is determined by a fit to a stretched decaying exponential:

\begin{equation}
\textrm{contrast(T}_1\textrm{)} = 1 - \textrm{A}  \textrm{e}^{-\tau \times \kappa(\textrm{T}_1)^{\epsilon}}
\end{equation}

\noindent where $\epsilon \approx 1$, the stretching term, is allowed to freely vary in the initial fit, but is fixed for the pixel-by-pixel GPU analysis. 

Rabi frequency is determined by a fit to a decaying sinusoid with functional form

\begin{equation}
\textrm{contrast(Rabi)} = 1 - \frac{\textrm{A}}{2} (1-\textrm{cos(}\tau \times 2 \pi \textrm{f)} e^{-\tau \times \kappa(\textrm{Rabi})}\textrm{)}
\end{equation}

\noindent with f being the Rabi frequency (in \SI{}{\hertz}). 

T$_2^*$ from Ramsey is measured with the applied MW signal detuned by $\approx$\SI{5}{\mega\hertz} from the NV center frequency: accordingly the relevant functional form for fitting is multiple overlapped decaying sinusoids corresponding to this detuning:

\begin{equation}
\begin{aligned}
\textrm{visibility(Ramsey)} = \textrm{[A}_\textrm{-1} (1-\textrm{sin(}\tau \times 2 \pi \textrm{(f-}\delta\textrm{))} + \textrm{A}_\textrm{0} (1-\textrm{sin(}\tau \times 2 \pi \textrm{(f))}\\
+ \textrm{A}_\textrm{+1} (1-\textrm{sin(}\tau \times 2 \pi \textrm{(f+}\delta\textrm{))}\textrm{]} e^{-\tau \times \kappa(\textrm{Ramsey})}
\end{aligned}
\end{equation}

\noindent where $\delta$ is the relevant NV hyperfine splitting for 14N (isotope used in diamond sample studied in present work), and f$\approx$\SI{5}{\mega\hertz} is the detuning. Note that for an NV diamond sample with 15N isotopic purification, the Ramsey visibility measurement has two detuned resonances instead of three. In the above fits, A or A$_i$ are freely varying amplitude parameters and $\kappa$ are the decay constants of interest. 

CW-ODMR is slightly more complicated: first the data is averaged together into smaller frames; then a peak-fitting algorithm is used to determine the 3 hyperfine locations for each individual NV resonance class. Finally, the location of these peaks is used to seed the following fit function:

\begin{equation}
\begin{aligned}
\textrm{contrast(CW-ODMR)} = 1-\textrm{A} (\textit{L}\textrm{(f}_i\textrm{-}\delta \textrm{,} \Gamma\textrm{)} + \textit{L}\textrm{(f}_i\textrm{,} \Gamma\textrm{)} + \textit{L}\textrm{(f}_i\textrm{+}\delta \textrm{,} \Gamma\textrm{))}
\end{aligned}
\end{equation}

\noindent where A is once again a free amplitude parameter, $\delta$ is the hyperfine splitting, $\Gamma$ is the CW-ODMR linewidth, f$_i$ is the center frequency of the NV resonance class, and $\textit{L}$(f,$\Gamma$) is a Lorentzian. This fitting is repeated per NV subclass for a total of eight sequential fits.

GPU enabled parallel computing resources allow data fits on the time scale of minutes. In the present work, GPUfit is employed: a Levenberg-Marquardt based non-linear least squares fitting package that utilizes Nvidia's CUDA toolset to enact computation on the GPU.\cite{GPUfit} The Levenberg-Marquardt algorithm has strict limitations, including limited robustness that can cause convergence to incorrect parameters if the initial estimate parameters are far away from the global solution to the least squares problem. For this reason, seed parameters are found using more advanced scipy least squares methods on averaged regions of the initial dataset, corresponding to a (NxN) dicing of the X-Y plane so that each seed value is a close approximate of the final solution before beginning LM fitting, with $1 \leq N \leq 10$. This approach reduces the problem to a small number of serially-calculated best fits ($\approx 100$) using more reliable fitting methods, before moving to the GPU to process the finer details.

\begin{figure}[htbp]
\centering
\fbox{\includegraphics[width=1\linewidth]{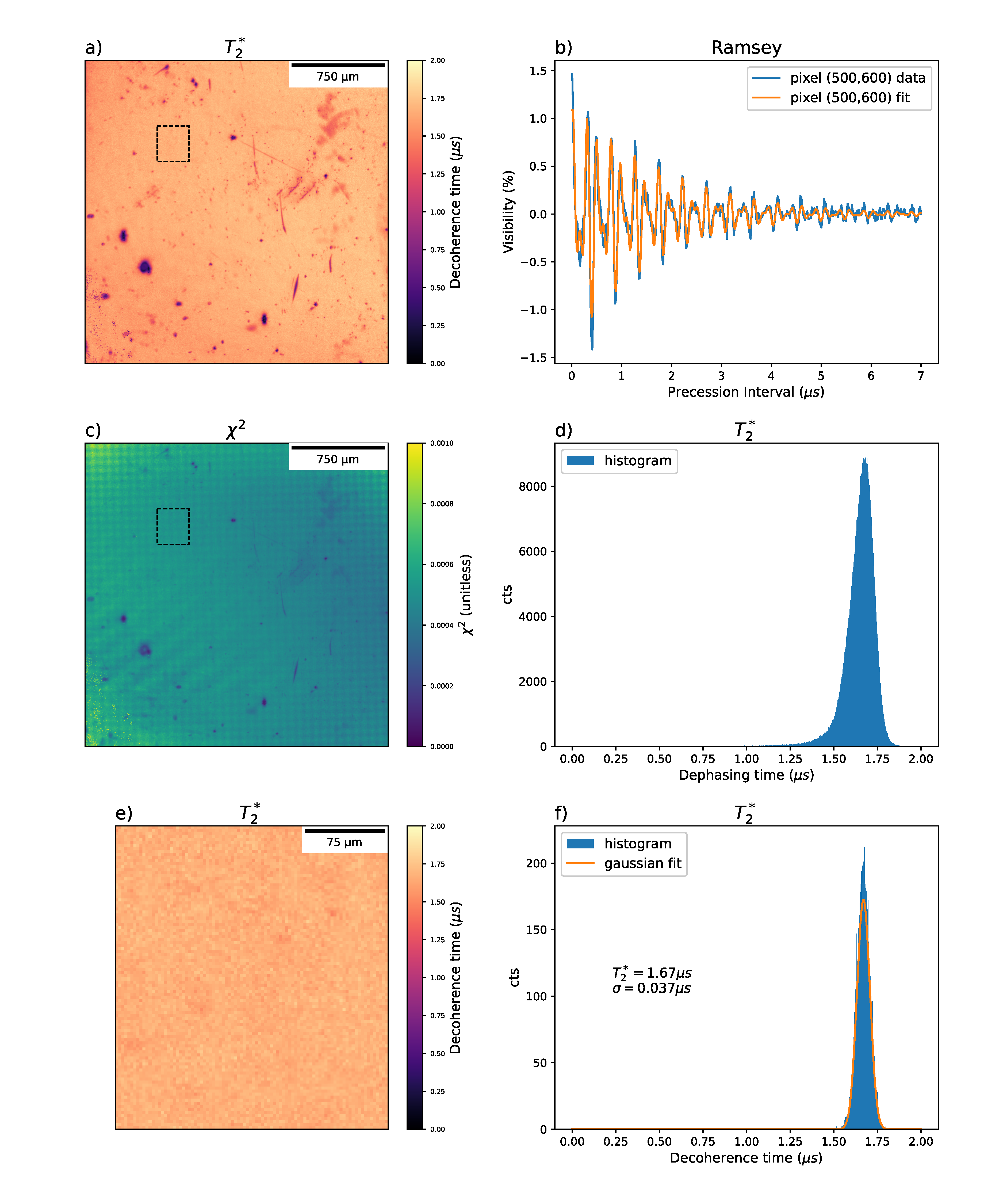}}
\caption{Evaluation methods for Ramsey fitting dataset. a) Map of NV diamond sample T$_2^*$ values, box corresponds to homogeneous (low strain) sub-region selected for further analysis. b) Example single-pixel Ramsey data and corresponding fit. Fit's decay envelope corresponds to pixel value used in T$_2^*$ map. c) $\chi^2$ map of associated fits in same field-of-view as T$_2^*$ map in a). d) Histogram of T$_2^*$ values from a), broadened asymmetrically by regions of strain and surface contaminants. e) Map of T$_2^*$ values for boxed sub-region in a). f) Histogram of T$_2^*$ values from e), and corresponding Gaussian fit giving median T$_2^*=1.67\mu s$ and standard deviation $\sigma=0.037\mu s$, corresponding to a confidence interval of about 2.2\%. Similar techniques are used to evaluate convergence in other imaged modalities.}
\label{fig:diagnostic}
\end{figure}

Several tools exist to analyze fit appropriateness in bulk. First, semi-random inspection of several distributed data-sets is used as a proxy for the overall accuracy of the ensemble; in particular, the aim is to choose extremal data-sets, e.g., from the edges of the sample or from the region of highest and lowest intensity, in order to confirm convergence under many conditions. An example of a single Ramsey data set is provided in Fig. \ref{fig:diagnostic} showing reasonable agreement between fit and a single-pixel data set. Additionally, inspection is performed of example regions of minimal strain to evaluate the convergence of fit parameters. This uses the random noise of the data set to approximate a Gaussian noise on the input data, and provides an estimation for the variance of the fit fidelity. Finally the chi squared outputs of fits can be imaged and provide insight into regions of good and poor fitting. Some degree of variance is expected between regions of high and low signal-to-noise ratio (SnR); however, if the fit function and seed parameters are well chosen, the $\chi^2$ will stay below a characteristically low value, given the number of free parameters and the data set size. An example of this process is demonstrated in Fig. \ref{fig:diagnostic}, where the $\chi^2$ map shows several interesting features. Notably, the edges of the map, where optical transmission and collection efficiency is poor, have high-$\chi^2$ values, which is expected for lower laser illumination and NV PL densities, corresponding to lower measurement SnR. Second, the regions of fast NV dephasing (low-T$_2^*$ value) have much lower-$\chi^2$ values. Although this result appears counter-intuitive, strong dephasing results in visibility $\approx$ 0 for most of the data set, leading to accurate fitting and uniformly low-$\chi^2$ values in these regions. Finally there is a visible grid pattern underlining the data set, which is a result of readout noise in the CMOS camera, resolved by sensitive fitting techniques.

\section{Acid Cleaning}

Samples are cleaned with a bench-top safe perchlorate etch process.\cite{AcidWash} Perchlorate, nitric, and sulfuric acids are mixed in a 5mL:5mL:5mL ratio solution. Diamond samples are added to this solution, and then the beaker is heated to 200$\degree$C in order to boil off both the nitric and nerchlorate acids. Sulfuric acid is still liquid at this temperature, and remains in the flask as the final product (as well as any etched contaminants). The fumes from the boiling process are passed through a condenser unit into a flask with DI water in order to capture the boiled acids for disposal. The post-condensed gas is then passed through a gas wash bottle in order to remove any trace remaining acid vapors. During setup, care must be taken that the total DI water used in the gas wash bottle and condenser flask do not exceed the total volume of the condenser flask, as water from the former can be ingested into the latter when the heat is turned off. 

\begin{figure}[htbp]
\centering
\fbox{\includegraphics[width=1\linewidth]{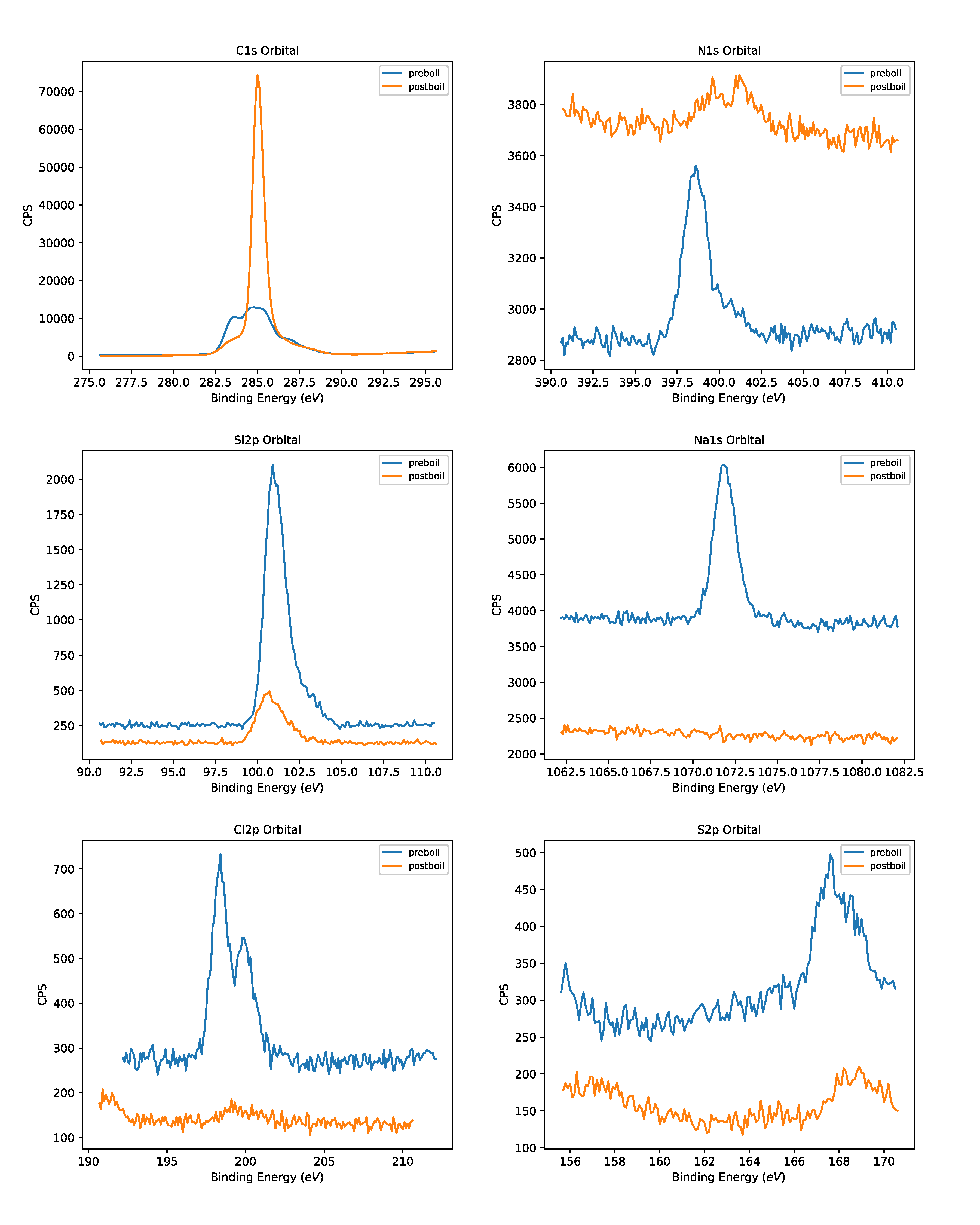}}
\caption{XPS results before and after perchlorate etch for several atomic species. Silicon and several common species decreased with etching, and Carbon 1s orbitals are significantly homogenized. Surface N also appears to be strongly suppressed.}
\label{fig:PerchlorateEtch}
\end{figure}

The result of the perchlorate etch process are diamond samples with reduced surface silicon contaminants, and all but eliminated Na and Cl, as confirmed by X-ray photoelectron spectroscopy (XPS), see Fig. \ref{fig:PerchlorateEtch}. Additionally, the process removes graphitic carbon, which can be detected by the shift in 1s-shell surface carbon. 

\bibliography{sample}